\documentclass[runningheads]{llncs}
\usepackage{graphicx}
\usepackage{multirow}
\usepackage{ragged2e}
\usepackage{tabularx,booktabs}
\usepackage{cite}
\usepackage{amsmath,amssymb,amsfonts}
\begin{document}
\title{Targeted attacks on the world air transportation network: impact on its regional structure}
\titlerunning{Robustness analysis of the world air transportation network}
\author{Issa Moussa Diop\inst{1}\orcidID{0000-0001-5624-319X} \and
Chantal Cherifi\inst{2}\orcidID{0000-0003-1619-7998} \and
Cherif Diallo\inst{1}\orcidID{0000-0001-6606-7337} \and
Hocine Cherifi\inst{3}\orcidID{0000-0001-9124-4921}}
\authorrunning{I. M. Diop et al.}
\institute{LACCA Lab, Gaston Berger University, Saint-Louis PB 234, Senegal \email{\{diop.issa-moussa,cherif.diallo\}@ugb.edu.sn}\and
DISP Lab, University of Lyon 2, Lyon, France
\email{chantal.bonnercherifi@univ-lyon2.fr}\\
 \and
LIB EA 7534, University of Burgundy Franche-Comté, 21078 Dijon, France\\
\email{hocine.cherifi@u-bourgogne.fr}}
\maketitle              % typeset the header of the contribution
\begin{abstract}
Perturbations of the air transport network have a tremendous impact on many sectors of activity. Therefore, a better understanding of its robustness to targeted attacks is essential. 
The literature reports numerous investigations at different levels (world, regional, airline) considering various targeted attack strategies. However, few works consider the mesoscopic organization of the network. To fill this gap, we rely on the component structure recently introduced in the network literature. Indeed, the world air transportation network possesses seven local components capturing the regional flights in localized areas. Its global component, distributed worldwide, capture the interregional routes. We investigate the impact of two influential attacks (Degree, Betweenness) on the world air transportation network at the regional and inter-regional levels. Results show that the seven regions are isolated one after the other from the world air transportation network. Additionally, although the Betweenness attack effectively splits the network, its impact on regional routes is less pronounced.

%The perturbation of the air transport network can cause an enormous impact on different sectors of activity. Then, the robustness of the world air transportation network is widely studied in the network literature. These investigations don't take into account the community structure of this network. This paper analyzes the robustness of the world air transportation network from a new perspective. Indeed, the large local components of the world air transportation network correspond to seven geographical areas. Then, it captures better the regional flights. In contrast, the global component distributed worldwide captures better the interregional flights. Then, removing an airport from the world has two consequences, one on regional flights and the other on interregional flights. The Degree and Betweenness centrality measures are the targeted attack. The giant component is the robustness metric. Results show that the world air transportation network is more resilient to the attack based on the Degree. Moreover, the local components are isolated as the world air transportation network is attacked. Once isolated, it is more challenging to travel within the LCC of the components, for the attacks are based on the Degree.

\end{abstract}
\section{Introduction}
Every day, millions of passengers and frets transit by air transport. This infrastructure is essential in people's lives for economic, social, and health purposes. A disturbance can have significant consequences in different sectors of activity. Therefore, it is vital to study its disruption to limit the damage. The complex network paradigm attempts to provide solutions in this sense. Indeed, representing the airports as nodes and the flights between two airports as links, the robustness of the air transportation network has been extensively studied. One uses a random attack to account for accidental service disruption in an airport. In this case, one evaluates the impact of removing nodes at random on the network topology. Robustness studies also consider targeted attacks. In this case, the goal is to elaborate a node removal strategy to damage as much as possible the network. Besides, one can distinguish studies either linked to geographical areas (worldwide, regional, national) or airlines in the literature. To contextualize our work, we briefly describe important related works. For more details about the robustness of the air transportation network, the reader can refer to \cite{lordan2014study, sun2021robustness}.

%----------------------------paper1 2----------------
In \cite{sun2017robustness}, the authors conduct an extensive analysis of the robustness of the world air transportation network. They use six centrality measures (Betweenness, Closeness, Eigenvector, Bonacich, and Damage) to quantify the importance of airports. They consider two attack strategies. A strong attack removes the nodes in descending order of the centrality measure, while a weak attack uses the inverse order. They perform the experiments on three networks: 1) the unweighted network, 2) the network weighted by the number of passengers, 3) the network weighted by the inverse distance between airports. They compare three robustness metrics when removing a node: 1) the size of the giant component, 2) the number of survived links, 3) the number of unaffected passengers with rerouting. 
Results show that the size of the giant component is not appropriate for evaluating weak attacks' robustness. Furthermore, it overestimates the robustness of strong attacks. In comparison, survived links fix these two anomalies. In addition, according to the unaffected passenger with rerouting, Degree centrality is the most effective attack when 7\% of nodes are disconnected. Bonacich takes the lead when more than 7\% are disconnected. The robustness is sensitive to the type of weight of the network.
%---------------------paper 2---------------------
In \cite{lordan2019core}, the authors investigate the robustness of the seven regional unweighted networks defined by the OAG (Africa, Asia, Europe, Latin America, Middle East, North America, and South Pacific). They perform strong attacks based on Degree, Betweenness, and Damage. The size of the giant component evaluates the robustness. Results show that Damage is the most effective attack for a small fraction of removed nodes. However, Betweenness performs better when the number of removed nodes grows. Results also show evidence that differences across regions are related to the size of the value of the k-core. Areas with a large core of densely connected airports, such as Europe and the Middle East, are more resilient than regions with a smaller core.
%-------------------paper3 4--------------------------
In \cite{zhang2021analysis}, the authors investigate the robustness and weighted (number of flights per week) and unweighted Belt and Road region network. Strong attacks on the unweighted network use Degree, Betweenness, Closeness, and Eigenvector centrality. For the weighted network, they investigate strong attacks based on recursive power and recursive centrality. They test four robustness metrics: clustering coefficient, average shortest path length, graph diversity, and global efficiency. Results show that the Betweenness centrality is the least robust for the clustering coefficient, average shortest path, and global efficiency for the unweighted network. The Degree is less resilient for the graph diversity. Recursive power is more sensitive to all the network properties for the weighted network. By comparing the weighted and unweighted network's robustness, no targeted attack dominates entirely. Indeed, the Betweenness centrality is less robust for the average shortest path and the graph diversity. The Degree is sensitive to graph diversity. Finally, the recursive power is less resilient for the clustering coefficient.
%----------------------paper 4-----------------
In \cite{lordan2016robustness}, the authors investigate the topology and robustness of unweighted networks of airlines. They study 10 Full-Service Carriers belonging to three airline alliances (Star Alliance, OneWorld, and SkyTeam) and 3 Low-Cost Carriers (Ryanair, EasyJet, and Southwest Airlines). Random and targeted strong attacks based on Degree and Betweenness centrality are performed. The size of the giant component is the robustness metric. Low-Cost Carriers are more resilient than Full-Service Carriers to random attacks, but the difference is not significant. The attack based on the Degree is more resilient than the attack based on the Betweenness. But the difference is small. The hybrid model airlines tend to be more robust than the Low-Cost Carriers airlines, which resist better targeted attacks than the Full-Service Carriers airlines.

Our work departs from previous studies. It concerns the interactions between the unweighted world air transportation network and its regional components. In earlier work, we introduced a network decomposition called the component structure of a network \cite{diop2021revealing}. It decomposes a network into its local components and global components. Local components are localized dense areas of the original network. The links joining the local components and their associated nodes form the global components. Based on this representation, air world transportation comprises several regional components corresponding to natural geographic and cultural areas. The inter-regional components reveal the main airports and routes between these regions. Based on this representation, we explore the impact of targeted attacks on the world air transportation network on its regional and inter-regional components. We consider Degree and Betweenness centrality measures to remove nodes in descending order. These experiments give new insight into the interactions between international and regional routes exposed to disruption. 

The rest of the paper is organized as follows. Section 2 introduces the background. Section 3 presents the data and methods. Section 4 reports the main findings of our analysis. Section 5 discusses the results and then gives conclusions.
\section{Background}
\subsection{Component structure}
The density of real-world networks is generally not uniform. One usually captures this phenomenon using two mesoscopic features: 1) the community structure, 2) the core-periphery structure. Although there is no consensus on a universal definition of these representations, they share the fact that the network contains groups of nodes tightly connected called core or communities. They are supposed to be loosely related to other groups when considering the community structure. Peripherical nodes sharing few connections surround these core groups in the multi-core-periphery structure. The component structure builds in these representations. It splits the networks into dense groups and their interactions. One obtains the local components by isolating the dense parts of the networks. Links and nodes connecting the local components form the global components. 
To build the component structure one proceeds as follows:
\begin{enumerate}
    \item Uncover the dense parts of the network.
    \item  Remove the links between the dense parts to extract the local components.
    \item  Remove the links within the dense parts and the subsequently isolated nodes to extract the global components.
\end{enumerate}
Note that this representation is redundant. Indeed, a node can simultaneously belong to a local and a global component. One can use community detection or multi-core-periphery algorithms to extract the dense parts of the network. We consider an approach based on the community structure to extract the components in this work. Fig.~\ref{fig:method}.A describes the extraction process of the component structure. In this example, one uses a non-overlapping community detection algorithm to extract the dense parts of the network. Then, we form the local components by removing the inter-community links. Removing the intra-community links and the isolated nodes extracts the global components.

\subsection{Targeted Attack}
%One can consider two types of attack in a network: random and targeted attacks. A random attack consists in removing a node randomly. It allows for studying the robustness of the network when a failure occurs. In contrast,%
Targeted attacks aim to remove the most vital nodes for network connectivity\cite{chakraborty2016immunization}. Centrality measures generally describe the importance of nodes \cite{ibnoulouafi2018m}. In a strong attack strategy, one removes nodes in the network in descending order of magnitude of the chosen centrality. This work uses the most popular measures: Degree and Betweenness.

%\subsubsection{Degree centrality}
\textbf{Degree centrality} of a node is the number of its first-order neighbors. Given a graph $G(V, E)$, such as V is the set of nodes and E the set of links, the Degree $k_{i}$ of node $i$ is defined as:
$$  
    k_{i} = \sum_{j \in V, i\ne j}{a_{ij}}
$$
$a_{ij}$ is an element of the binary adjacency matrix of $G$ such as $a_{ij}=1$ if $i$ and $j$ are connected, otherwise, $a_{ij}=0$. \\
%\subsubsection{Betweenness centrality}
\textbf{Betweenness centrality} of a node is the fraction of the shortest path passing through it. When it is normalized, the Betweenness of the node $i$ is defined as:
$$
    b(i) = \frac{2}{(n-1)(n-2)}\sum_{i \ne j} \frac{\sigma_{jk(i)}}{\sigma_{jk}}
$$
$\sigma_{jk}$ is the number of the shortest path between $j$ and $k$. $\sigma_{jk}(i)$ is the number of the shortest path from $j$ to $k$ passing in $i$.
\subsection{Evaluation measures}

The size of the giant component is the most popular metric to assess the robustness of a network. It refers to the size of the largest set of interconnected nodes when the network brakes into several parts due to node removal. The higher the size of the giant component, the most resilient the network is to the attack.

%The size of giant component is the most used metric to assess the robustness of a network. It refers to the size of the largest set of the interconnected nodes when the network is disrupted. The attack that decreases slowly the size of giant component is the most resilient.
\section{Data and Methods}
\subsection{Data}
We consider an unweighted and undirected network originating from FlightAware\cite{flightaware}. Flight information has been collected for six days (between May 17, 2018, and May 22, 2018). Nodes represent airports, and links represent direct flights between airports during the period. Table~\ref{table:main communities} reports its basic topological properties.

\subsection{Methods}
Our goal is to evaluate the impact of a targeted attack on the world route network on its regional and inter-regional constituents. Therefore, once the component structure is extracted, the robustness evaluation process proceeds as follows:
\begin{enumerate}
    \item Disconnect the node from the world air transportation network according to an attack strategy.
    \item Disconnect the same node from its local component. 
    \item Disconnect the same node from the global component if it also belongs. 
    \item Extract the giant component from the world air transportation network.
    \item Extract the giant component from the concerned local component.
    \item Extract the giant component from the global component.  
\end{enumerate}

\begin{figure}
\includegraphics[width=\textwidth]{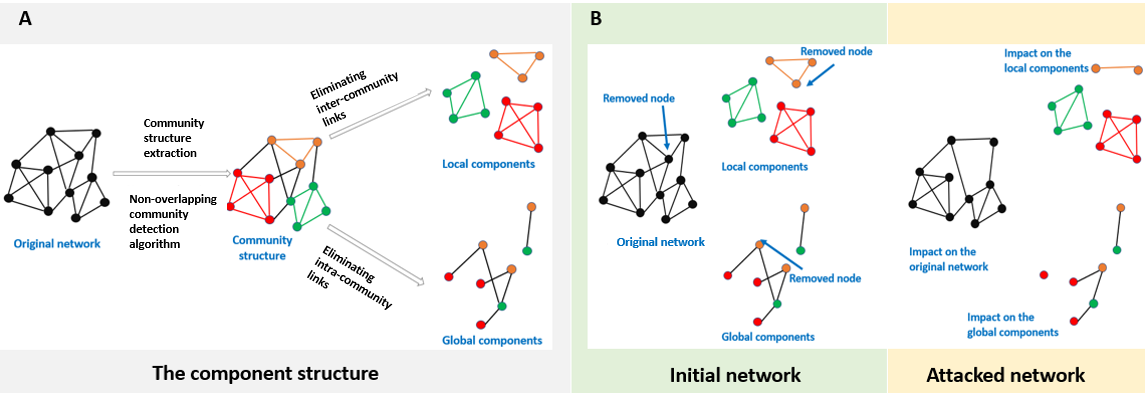}
\caption{(A) Process to uncover the component structure. The community structure is used for example, to extract the dense parts (B) Attack on the network.} \label{fig:method}
\end{figure}

This approach allows us to visualize the impact of removing a critical airport in the world air transportation network on the robustness of the regional and inter-regional networks. Fig.~\ref{fig:method}.B reports a toy example illustrating an attack on an airport of the world network and the disruptions induced in the local and global components of the network.  

%------------------------------------------------
%----------------------------------------------------
\section{Experimental results}
\subsection{Component structure}
We rely on the Louvain community detection algorithm to uncover the dense parts of the network. Indeed, communities are tightly connected nodes. It reveals 27 communities. Therefore, we consider that there are 27 local components. There are seven large and twenty small components localized in various geographical areas. The largest local components cover the following regions: 1) North America-Caribbean, 2) Europe, 3) East and Southeast Asia, 4) Africa-Middle East-Southern Asia, 5) Oceania, 6) South America, 7) Russia-Central Asia-Transcaucasia. Note that this subdivision is different than the partition proposed by the OAG (Official Airline Guide) used in \cite{lordan2019core}.

\begin{figure}
\includegraphics[width=\textwidth]{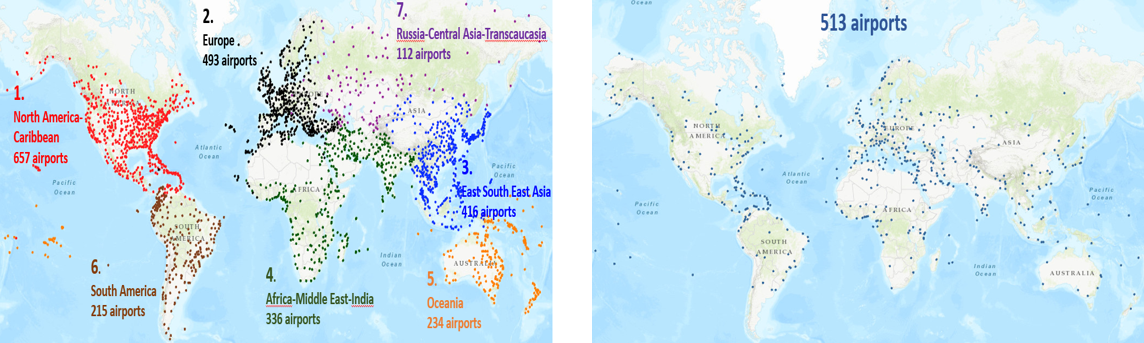}
\caption{Left figure:The airports in the seven large local components. Each color represent a component. Right figure: The largest global component.} 
\label{fig:large_comp}
\end{figure}

There are fifteen global components. The largest one regroups more than 96\% of the airports, and it is distributed over the world \cite{diop2021revealing}. Fig.~\ref{fig:large_comp} represents the airports included in the largest local and global components. We restrict our attention to the seven largest local components and the main global component in the following robustness analysis. Table~\ref{table:main communities} reports their basic topological properties.
%----------------------------------------------
%------------------------------
\begin{table*}[t!]

\caption{Basic topological properties of the world air transportation network, the 7 large local components, and the largest global component. $N$ is the network size. $|E|$ is the number of edges. $diam$ is the diameter. $l$ is the average shortest path length. $\nu$ is the density. $\zeta$ is the transitivity also called global clustering coefficient. $k_{nn}(k)$ is the assortativity also called Degree correlation coefficient. $\eta$ is the hub dominance.}

\label{table:main communities}
\begin{tabular}{|l|*{8}{c|}}\hline

  Components &  $N$ & $|E|$  & $diam$ &  $l$ & $\nu$ & $\zeta$  & $k_{nn}(k)$ & $\eta$  \\
\hline
World air transportation network  &2734   & 16665    & 12  & 3,86  &  0,004  & 0,26     & -0,05 & 0.09  \\  \hline \hline

North America-Caribbean &  657 & 3828 & 7 &  2.88 & 0.018 &  0.28 & -0.325 & 0.29   \\ \hline

Europe &  493  & 5181  & 6 & 2.58  & 0.042 & 0.32  & -0.2 & 0.33  \\ \hline

East and Southeast Asia &  416  & 2495  &  7 &  2.9 & 0.029  &  0.34   & -0.22 & 0.32     \\ \hline

Africa-Middle East-India &  337 & 1197  &  6  &  3.25 & 0.021   & 0.28  & -0.15 & 0.23  \\ \hline

Oceania &  234 & 464  & 9 & 3.5 & 0.017   & 0.18    & -0.22  & 0.24    \\\hline

South America &  215 & 527 & 6  & 3.15 & 0.023 & 0.23 & -0.35 & 0.22      \\ \hline

Russia-Central Asia-Transcaucasia &  112 & 427 & 4  & 2.23 & 0.07   & 0.23    & -0.39 &  0.73    \\\hline \hline
Large global component & 513 &  2194  & 8  & 3.28 & 0.017  &  0.13  & -0.25 & 0.2  \\ \hline

\end{tabular}
\end{table*}
%-------------------------------------------------------
\subsection{Degree targeted attack}
\begin{figure}
\includegraphics[width=\textwidth]{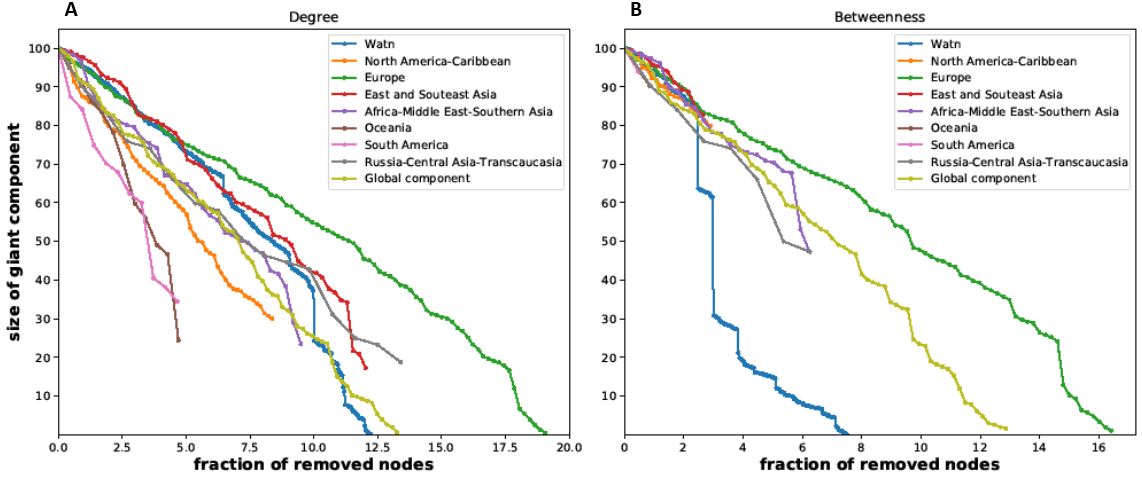}
\caption{The size of the giant component of the world air transportation network, and the large components as a function of removed nodes under targeted attacks on the world air transportation network.} \label{fig:target_attack}
\end{figure}

Fig.~\ref{fig:target_attack}.A illustrates the evolution of the LCC with the fraction of top Degree nodes removed from the world air transportation network in descending order of the airports' Degree centrality. We also plot the induced evolution of the LCC of the various large components. All the curves exhibit similar behavior. As the fraction of nodes removed from the world air transportation network grows, the LCC of the components decreases almost linearly. Overall, the European component is the most resilient. The East and Southeast Asia components and the World Air Transportation Network follow. They behave similarly when the fraction of removed nodes is below 5\%. Beyond this value, the gap between the evolution of their LCC with the European component widens. Africa-Middle East-Southern, Russia-Central Asia-Transcaucasia, and the largest global components form a group of overlapping curves when the fraction of removed nodes is below 7.5\%. Beyond this value, they diverge. The North America-Caribbean component follows. Finally, Oceania and South America components are the most sensitive to the attack. Indeed, their LCC breaks down to 33\% and 28\% respectively, after removing around 4.7\% of the airports. 
%------------------décroissance des composantes:--------------
A more detailed observation shows two types of behavior. In the first case, the LCC exhibits piecewise linear variations. It concerns South America, Oceania, and Russia-Central Asia-Transcaucasia. Removing a particular node produces a sharp drop in the LCC because an entire subnetwork broke away from the primary component, causing considerable damage.
In contrast, in the other components and the world air transportation network, the size of the LCC varies almost linearly with the fraction of removed nodes. The same proportion of nodes leaves the LCC, whatever the node removed in this situation.

We now look at when the components become isolated from the world air transportation network. We identify the critical airport disconnecting the component. We explore the topological properties of the remaining regional network to evaluate the impact on the regional traffic. Table~\ref{table:remain_comp} displays the basic topological properties of the components after isolation. %Table~\ref{table:remain_comp} displays the basic topological properties at the same time as the component is isolated.
%-----------Isolation, 1) pourcentage réseau mondial , 2) nombre d'aéroports supprimé ) 3 aeroport critique, 3 proprietes de la nouvelle composante
Oceania is the first component to separate from the world air transportation network. It occurs after removing 9\% of the top Degree nodes from the world air transportation network. The proportion of top Degree nodes of this component subtracted before the split equals 4.7\%. Christchurch, in New Zealand, is the last top-Degree airport removed, provoking the separation. Only 24\% of the airports remain in the LCC of this component. They are all in Australia. All the other countries are unreachable. Although the LCC is denser than the initial component, traveling in this region becomes more challenging. Indeed, Table~\ref{table:remain_comp} shows that the diameter and the average shortest path increase, and the transitivity decrease.
%------------------------------------------------------------
The North America-Caribbean local component is the second to break away after removing 9.7\% of the top Degree nodes from the world air transportation network. When it happens, 8.4\% top Degree airports in this component have already been removed. Winnipeg /James A Richardson Airport in Canada is the last airport connecting the region to the world before the split. In contrast to Oceania, airports in the LCC are distributed in the different countries of the component. However, Alaska is isolated, and in Canada, there are only a few airports in Quebec and Ontario. The remaining LCC contains 30\% of the airports of the initial component. Table~\ref{table:remain_comp} shows that travels in the region become uneasy. Indeed, eight hops on average and 21 at maximum are required to join any two airports. In addition, density and transitivity decrease.
%---------------------------------------------
\begin{table*}[t!]
\caption{Basic topological properties of the 7 large isolated local components after the attack based on the Degree. $LCC$ is the Largest Connected Component. $|E|$ is the number of edges. $diam$ is the diameter. $l$ is the average shortest path length. $\nu$ is the density. $\zeta$ is the transitivity also called global clustering coefficient. $k_{nn}(k)$ is the assortativity also called Degree correlation coefficient. }

\label{table:remain_comp}
\begin{tabular}{|l|*{8}{c|}}\hline

  Components &  $LCC$ & $|E|$  & $diam$ &  $l$ & $\nu$ & $\zeta$  & $k_{nn}(k)$ & $LCC$(\%) \\
\hline
North America-Caribbean &  196 & 301 & 21 & 7.22 & 0.015 &  0.21 & -0.14  &  29.8  \\ \hline

Europe &  136  & 177 & 13 & 5.75 & 0.019 & 0.07 & -0.13  & 27.5 \\ \hline

East and Southeast Asia &  72  & 104 & 11 & 4.51 & 0.04  &  0.18 & -0.15  & 17.2 \\ \hline

Africa-Middle East-India &  79 & 110  & 17 &  6.9 & 0.035 & 0.26  & -0.08 & 23.5 \\ \hline

Oceania & 57 & 65 & 15 & 5.1 & 0.04 & 0.12 & -0.2  & 24.3  \\\hline

South America &  74 & 98 & 13 & 5.7 & 0.036 & 0.21 & -0.43 & 34.4 \\ \hline

Russia-Central Asia-Transcaucasia &  21 & 28 & 6 & 2.85 & 0.13  & 0.08 & -0.11 & 18.7\\\hline

\end{tabular}
\end{table*}
%-----------------------------------------
Disconnecting 9.8\% of the airports in the global air network is sufficient to isolate the South America component. A proportion of 4.6\% top Degree nodes are removed from the component before separation. Salgado Filho Airport (Brazil) is the last link with the rest of the world before the break. The LCC contains 34\% of the airports of the initial component. Most of its airports are in Brazil and its neighboring countries. Argentina is unreachable. Even though it is denser than the initial component, the shortest path and the diameter reveal that this network doesn't facilitate the traffic (see Table~\ref{table:remain_comp}). Note that the transitivity doesn't fluctuate much.
%-----------------------------------------------------
The next component that becomes unreachable is East and Southeast Asia. It breaks away after removing almost 11\% of the top hubs from the overall network. Before the split, 12\% of the top Degree nodes of this component have been removed. Gimhae Airport in South Korea is the last airport reachable before separation. The LCC contains 17\% of the airports from the initial component. They are mainly in China and its satellite countries. Compared to the other isolated component, it is the easiest to travel, even though its transitivity is low.
%--------------------------------------------------------
One needs to eliminate 11.2\% of the high Degree airports of the world air network to isolate the Africa-Middle East-Southern Asia region. The targeted hubs include 9.5\% of the airports of this component. The Benazir Bhutto Airport in Pakistan is the last connected to other areas in the world. The LCC contains 23.5\% of the airports of the initial component. They are located mainly in West-Central Africa and India. Traveling into the LCC is difficult even though it is denser than the original component. Almost all the Middle East, South Africa, and East of Africa airports are unreachable. Moreover, a maximum of 17 hops are required to join any two airports, while on average, it is seven hops.
%-----------------------------------------------------------
Russia-Central Asia-Transcaucasia is the next component to separate from the world air transportation network. One needs to remove 11.4\% of the top hubs of the world, including 13.4\% of this component's hubs. The last removed airport, the Heydar Aliyev Airport, is in Azerbaijan. Only 18.7\% of the airports initially in the component remains in the LCC. Except for two airports located in Uzbekistan and Tajikistan, All the others are in Russia. Compared to the initial component, the topological properties do not change much. Traveling in what remains of the component is as easy.
%-----------------------------------------------------
Once Russia-Central Asia-Transcaucasia is isolated, all the airports remaining in the world air transportation are in Europe. It occurs after removing 19\% of the top Degree nodes of the Europe component. The LCC contains 27.5\% of the airports of the original component located in various countries. However, countries like Spain, Norway, Sweden, and Finland are almost unreachable. It is more difficult to travel within the LCC than in the initial component. Indeed, the diameter and the average shortest path values double. In addition, the density and the transitivity decrease significantly. Finally, there are no more routes in the world air transportation network after removing 12.2\% of its most influential airports.
%---------------------------------------------------------
\subsection{Betweenness targeted attack}
Fig.~\ref{fig:target_attack}.B displays the variation of the LLC as a function of the fraction of top Betweenness centrality airports removed from the world air transportation network. We also plot the corresponding evolution for the components. Removing up to 3\% of top airports does not significantly differ in the various components. Indeed, the curves are very close. Nevertheless, Africa-Middle East-Southern Asia, East and Southeast Asia, and Europe are slightly more resilient. Above this value, Europe appears clearly as the most robust component. One can still distinguish the two types of behavior observed in the previous experiment. The LCC varies almost linearly in the Europe component and to a lesser extent in the global component. Piecewise linear variation with sharp drops characterizes the others and, more particularly, the World air transportation network.

%-----------------------------------------------------------
South America is the first isolated area after removing 1.6\% of the world's major airports. The last link with the World is Comodoro A M Benítez airport in Chili. Only two airports in this region are in the top 1.6\% world airports (Guarulhos-Governador, André F Montoro in Brazil, and Comodoro A M Benítez airport in Chili). The LCC contains 93.5\% of the airports of the initial component scattered in all the countries. The topological properties of the LCC do not change significantly as compared to the initial component, and the regional traffic keeps its efficiency.
%-------------------------------------------------
Removing 2.1\% of the top airports disconnect Oceania from the world transportation network. Perth airport is the last connection before isolation. Only one airport (Sydney K Smith) before Perth has been targeted before isolation. The LCC retains 90.2\% of the airports from the initial component across all its countries. The topological properties do not change much, and the internal travels in this component are not much disturbed.
%Removed 2.1\% of the influential airports of the world are isolated in the Oceania region. The last connected airport to the world is the Perth airport. Only one airport(Sydney K Smith) before Perth is targeted. This component conserves 94.8\% of these airports. The later are distributed between the countries of this component. The topological properties have not changed much, and the internal travel in this component is not much disturbed, like in South America. 
%At once, the world air transportation and the global component reduces to 86.1\% and 46.6\% respectively. 
%--------------------------------------------
East and Southeast Asia is the next component to become unreachable from the other parts of the world after removing 2.3\% of the world's most important airports. Among them, there are 2.6\% of the airports in this region. The last connection is through the Kunming Changshui Airport in China. The LCC contains 82.4\% of the airports initially in the component. They are located in the various countries of this component. Compared to the initial component, the topological properties of the LCC are comparable. Therefore, the impact of the isolation on regional routes is limited.
%When 2.3\% of the world's most important airports are removed, the East and Southeast Asia component becomes unreachable. Among them, there are 2.6\% of the airports in this region. The last removed airport in this component is the Kunming Changshui Airport in China. Moreover, 84.4\% of the airports, located in the various countries of this component, are interconnected. Compared to the initial component's topological properties, one can say that the efficiency of the remained network decreases a bit. 
%The world air transportation network and the global component falls to 84.7\% and 41.5\% of these airports.
%--------------------------------
North America-Caribbean is the fourth isolated area. It becomes unreachable after removing 2.9\% of the top central world airports. Among these, there are the top 2.9\% of this component. Simón Bolívar Airport in Venezuela is the last liaison to the world before isolation. The LCC contains 79.6\% of the airports still interconnected. Note that the Alaska subregion is inaccessible. The LCC is a bit less efficient than the component. Indeed, it keeps some large hubs, except in Canada. The topological properties of the LCC indicate that it is more challenging to travel than in the initial network.
%The fourth isolated area is the North America-Caribbean. Indeed, it becomes unreachable when 2.9\% of the central airports are removed. The top Degree nodes that are removed in this component correspond to 2.9\%. The last removed among them is the Simón Bolívar Airport in Venezuela. Note that, there are some airports from Venezuela in this component. Therefore, 80\% of the airports are still interconnected. Note that the Alaska subregion is even completely inaccessible. This remained network is a bit less efficient than the component. Indeed, It conserves some largest hubs, except the hubs of Canada. The topological properties fluctuate a bit. 
%At present, the world air transportation network diminishes fast to 61.6\%. Whereas the global component decreases to 32\%. 
%-----------------------------------------------
The next isolated area is the Africa-Middle East-Southern Asia, when one removes 3.9\% of the central airports from the world network. Around 5.6\% of the airports from this component are the target of the attack. As for the Degree centrality, Benazir Bhutto Airport in Pakistan is the last removed. The LCC maintains 47.6\% of the airports in this region connected. Once again, the Middle East and the Horn of Africa are the most impacted. They are almost unreachable. Traveling within this region becomes difficult. Indeed, the diameter (14 hops) and the average shortest path (6 hops) of the LCC are two times higher. The density and the transitivity are comparable with the initial component values.
%The next isolated area is the Africa-Middle East-Southern Asia, when 3.9\% of the central airports are removed from the overall network. 5.6\% of the airports from this component are targeted by the attack. The Benazir Bhutto Airport, in Pakistan, is also the last removed, like for the Degree. Therefore, 47.6\% of the airports of this region are connected. The Middle East region is still the most impacted. This region is almost inaccessible. It is the same behavior in the Horn of Africa. This part is quasi unreachable. Then, The movement within this region becomes difficult. indeed, the diameter (14 hops) and the average shortest path(6 hops) have doubled. The density and the transitivity are practically the same. 
%The world air transportation network falls rapidly to 20.3\% and the global component to 19.6\%. 
%--------------------------------------------------
Russia-Central Asia-Transcaucasia is the next region to break away from the world network after removing the top 5\% central airports of the global air transportation network. The attack involves 6.2\% major airports of this component before the separation. They are mainly in Russia. The last removed airport is Krasnodar Pashkovsky Airport. The size of the LCC reduces to 47.3\% of the initial component. Comparing its topological properties to the initial component allows us to conclude that the impact on regional easiness to travel is limited. 
%Removed the 5\% of the central airports of the global air transportation network, makes Russia-Central Asia-Transcaucasia the 6th. Before the separation, 6.2\% central airport of this component are attacked. There are particularly located in Russia. The last removed airport is Krasnodar Pashkovsky  Airport. By the way, The size of the LCC, of this component diminished to  56.4\%. It is almost the same proportion of impacted as the Africa-Middle East-Southern Asia region. The topological properties of this network change a bit. Indeed, on average, the number of jumps to join two airports is the same as the component. The diameter is incremented of 1, and the density and the transitivity are quasi the same.  
%At the same time, 85\% of the airports of the global air networks become unavailable. The global component is almost break down. Only the European airports subsist. 
%----------------------------------------------------
After splitting with Russia-Central Asia-Transcaucasia, Europe preserves 61.8\% of its airports in the LCC.  They are scattered in all the countries of the initial component. Compared to the latter, the LCC is less dense. In addition, one requires more hops to join any two airports.
%%?????????The European region is the most resilient area. This remained network preserves 27.6\% of the airports distributed in all the  countries of the component. Compared to the latter, this network is less dense. In addition, more hops are required to join any two airports. It breaks down with the world air transportation network when 7.5\% of the central airports are removed, in which 81 are located in the European region.??????????
%---------------------------------------------------
\begin{table*}[t!]
\caption{Basic topological properties of the 7 large isolated local components after the attack based on the Betweenness. $LCC$ is the Largest Connected Component. $|E|$ is the number of edges. $diam$ is the diameter. $l$ is the average shortest path length. $\nu$ is the density. $\zeta$ is the transitivity also called global clustering coefficient. $k_{nn}(k)$ is the assortativity also called Degree correlation coefficient.}

\label{table:remain_bet_comp}
\begin{tabular}{|l|*{8}{c|}}\hline

  Components &  $LCC$ & $|E|$  & $diam$ &  $l$ & $\nu$ & $\zeta$  & $k_{nn}(k)$ & $LCC$(\%)  \\
\hline
North America-Caribbean &  523 & 1918 & 8 & 3.23 & 0.14 & 0.2 & -0.27 & 79.6\\ \hline

Europe &  305  & 1252 & 8 & 3.39 & 0.027 & 0.16 & 0.014 & 61.8\\ \hline

East and Southeast Asia &  343  & 1421 & 8 & 3.51 & 0.24 & 0.32 & -0.11 & 82.4 \\ \hline

Africa-Middle East-India &  160 & 290 & 14 & 5.29 & 0.022 & 0.26  & -0.25 &  47.6\\ \hline

Oceania & 211 & 383 & 9 & 3.6 & 0.017 & 0.17 & -0.21  &  90.2\\\hline

South America &  201 & 449 & 7 & 3.32 & 0.022 & 0.21 & -0.38 &     93.5 \\ \hline

Russia-Central Asia-Transcaucasia &  53 & 105 & 5 & 2.84 & 0.07  & 0.2 & -0.32 & 47.32 \\\hline

\end{tabular}
\end{table*}

\section{Discussion and Conclusion}
In this work, we leverage the component structure of the world air transportation network to analyze the impact of targeted attacks on the regional and inter-regional routes. We consider attacks based on two influential centrality measures. The first one, based on Degree, removes nodes according to their number of connections in descending order. The second one uses the proportion of shortest paths transiting through a node to rank the nodes. We perform the experiments on an undirected and unweighted network to focus on routes rather than flights or passengers.

%comparer l'efficacité des attaques.  Expliquer
Results show that whatever the attack on the world air transportation network, one can link the way the network disintegrates to the component structure. Indeed, components get isolated one after the other when removing a certain proportion of the top centrality nodes from the world air transportation network. The main differences between the two types of attack are the fraction of top removed nodes needed for a component to break away from the world network and the size of the remaining isolated component. Globally, one needs to target more nodes before isolation in the Degree attack than the Betweenness attack. Furthermore, the isolated components' size is smaller in the Degree of attack. Indeed, it prioritizes the top internal hubs in the components, reducing their size. One needs to remove many internal high Degree nodes before reaching the inter-regional airports tying the components together. In contrast, Betweenness centrality focuses on inter-regional airports, splitting earlier the components with minor damage to their internal structure. 
%Results show that the world air transportation is more sensitive to the attack based on the Betweenness centrality. This attack isolates the various local components quickly. One can see the Damage on the curves of the LLC concerning the world air transportation network, which decreases sharply. The traffic efficiency within their component remains unchanged. Indeed, the airports with the highest Betweenness play a bridge role. These airports connect various components. Consequently, interregional flights are more affected. In contrast, the regional flights are the most affected by the attack based on the Degree. The components are isolated gradually, and the LCC curve of the world air transportation network diminishes slightly. Then, the traffic becomes very challenged within their component.
% comparer l'ordre d'isolation des composantes
One after the other, South America, Oceania, East, Southeast Asia, and North America are isolated regions by the Betweenness attack.  Oceania, North America, South America, and East and Southeast Asia leave the world network in this order when targeting nodes by Degree centrality. Then Africa-Middle East-Southern Asia, followed by Russia-Central Asia-Transcaucasia separate from Europe in the two attack strategies. Note that economic and geographic ties connect these two groups of regions.
%As we can see, the component isolation is done by step for the two attack strategies. Then, Europe, Russia-Central Asia-Transcaucasia, and Africa-Middle East-Southern Asia are the last separate regions for the attack strategies. These two regions are connected to Europe by their position and economy. South America, Oceania, East, Southeast Asia, and North America are the first isolated for the Betweenness attack. Whereas, for the Degree, Oceania, North America, South America, and East and Southeast Asia are the first isolated. Indeed, these four regions are a bit remote.

% souligner l'interet de la representation en composante.permet d'evaluer l'impact au niveau regional%

The component structure allows us to study the robustness of the world air transportation network from a new perspective. Indeed, it is a mixture of subnetworks with various internal densities geographically well identified. Looking at its robustness through its component structure allows us to highlight the impact of targeted attacks in different world areas. These results will enable us to tailor the protection strategies to maintain inter-regional routes and minimize the effect of disruption at the regional level. Future work will also consider developing more effective attack strategies based on the component structure.

%The representation of the component structure allow us to perform the robustness of the world air transportation network from a new perspective. As we know, the world air transportation network presents a community structure. This feature has never been taken into account in the analysis of its robustness. With the component structure, this work attempt to fill this gap. 

%
%Perspectives
%However, all the new perspectives are explored in this paper. One can develop various strategy attacks based on the component structure. Moreover, the robustness of the local components and the global component can be exclusively studied. 

\bibliographystyle{jhep}
\bibliography{biblio}

\end{document}